\documentclass[9pt,twoside,twocolumn,a4paper]{article}
\usepackage{geometry}
\usepackage{graphicx}
\usepackage{subfig}
\usepackage{textcomp}
\usepackage{siunitx}
\usepackage{makeidx}
\usepackage{multicol}

\makeindex

\newcommand{\fig}[1]{Fig.\ref{#1}}

\title{Analysis of Dispersive Fourier Transform dataset using Dynamic Mode Decomposition: evidence of multiple vibrational modes, and their interplay in a three-soliton molecule.}
\author{Anastasiia Sheveleva$^1$, Sa\"{\i}d Hamdi$^{1,2}$, Aur\'{e}lien Coillet$^1$, Christophe Finot$^1$, Pierre Colman$^{1,*}$ \\
		$^1$Laboratoire Interdisciplinaire Carnot de Bourgogne, UMR 6303 CNRS, \\ Universit\'{e} de Bourgogne, 9 av. Alain Savary, 21000 Dijon, FRANCE \\
		$^2$Current affiliation: Menlo System GmbH, Bunsenstraße 5, 82152 Martinsried, GERMANY \\
		$^*$Corresponding author: pierre.colman@u-bourgogne.fr}
\date{}

\begin{document}
\maketitle

\par\textbf{ We demonstrate that the Dynamic Mode Decomposition technique can effectively reduce the amount of noise in Dispersive Fourier Transform dataset; and allow for finer quantitative analysis of the experimental data. We therefore were able to demonstrate that the oscillation pattern of a soliton molecule actually results from the interplay of several elementary vibration modes.}\\

Since its first demonstration, the Dispersive Fourier Transform (DFT) is a key technique to investigate experimentally ultrafast fiber ring laser cavities \cite{Jannson1983,Goda2013,Mahjoubfar2017,Wang2020,Godin2022}. As a convenient way to observe the pulse evolution round-trip after round-trip, this technique allows indeed a better comprehension of the nonlinear phenomena governing these systems. If the main features of fiber ring laser cavities are now well understood, it appears that weaker effects are also of importance because they can built-up to the point where their contribution to the dynamics becomes also determinant \cite{Jang2013, Hamdi2022}. Correspondingly, DFT experiments would now require either very long record time – so that the long-term dynamics caused by the weaker effects can fully develop; and more acute measurements – so that minute phenomena can be observed. Experimentally, the bottleneck would therefore come from the oscilloscope which records the DFT trace. Indeed, the electronic noise and the discretization granularity prevent recording the single shot spectra with utmost precision.\\ Using the example of a three-soliton molecule, we demonstrate in this letter that the Dynamic Mode Decomposition (DyMD) allows minimizing experimental noise without compromising the quality of the information. We show in particular that this technique is more efficient than any other curve smoothing techniques, for instance, the Savitzky-Golay filter \cite{Savitzky1964}.  Namely, we fully characterize the oscillations of a 3-Soliton Molecule (SM) and demonstrate that it comprises two nonlinear oscillators of different timescales that are coupled to each other. On a minor side, DyMD permits lossless compression of the DFT dataset by at least a 90 \% factor. 

\begin{figure}[!ht]
	\centering
	\includegraphics[width=\linewidth]{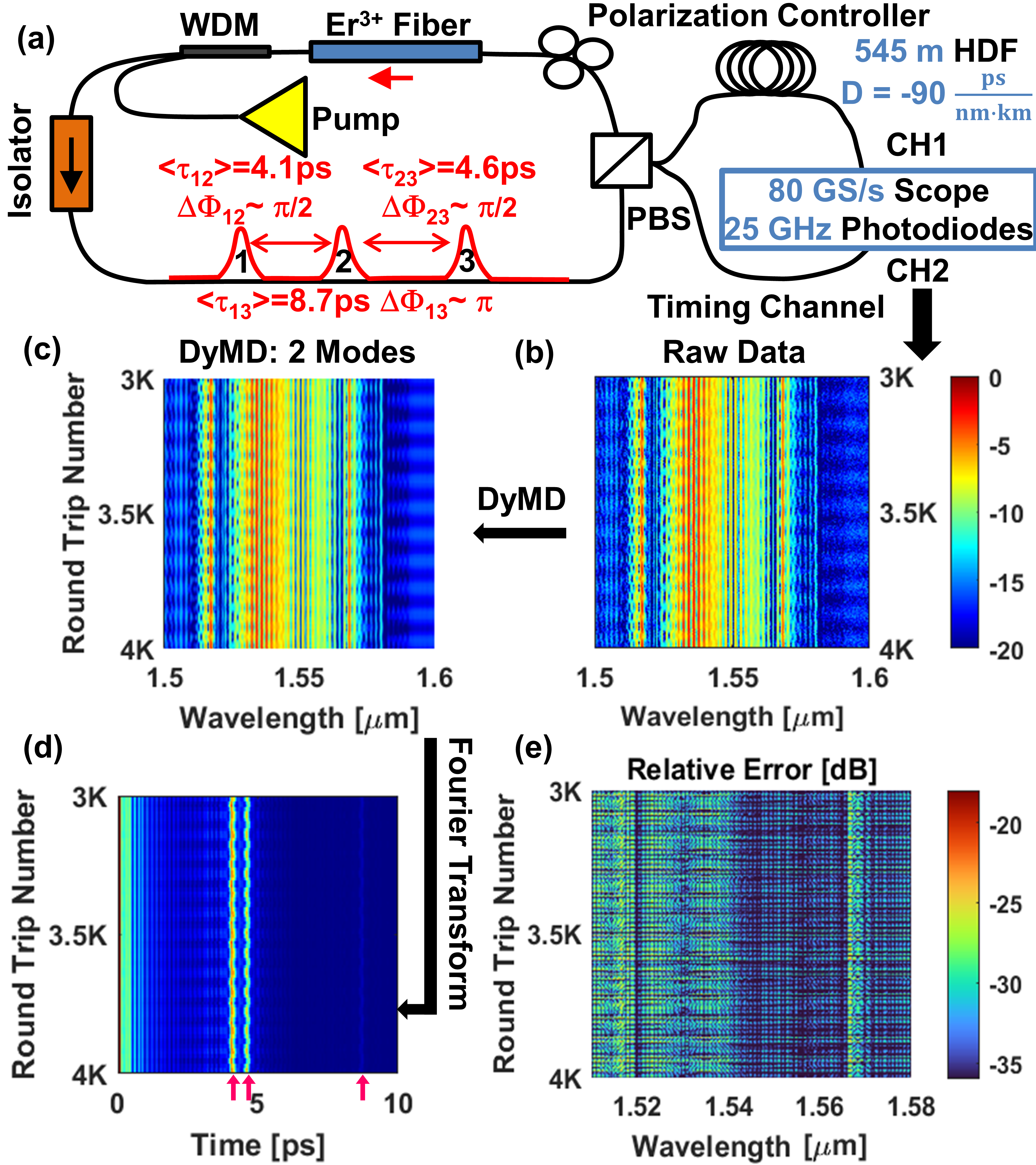}
	\caption{(a)  Experimental setup; PBS: polarization beam splitter; HDF: 545~m of Highly Dispersive Fiber; Pump: 980-nm laser diode; WDM: pump-signal multiplexer.  (b) Single-shot spectra. (c) DFT Spectra after 2-Mode reduction by Dynamic Mode Decomposition. (d) (logarithmic scale) Fourier Transform of (b) showing three characteristic inter-Soliton seperations of $4.1$~ps $4.6$~ps and $8.7$~ps (pink arrows). (e) Error of the DyMD in (b). -3~dB corresponds to 1 Bit of precision (hence -24~dB is 8-Bit precision). Subpanels (b-e) are represented in logarithmic scale.}
	\label{Fig:1}
\end{figure}

\begin{figure}[!ht]
	\centering
	\includegraphics[width=\linewidth]{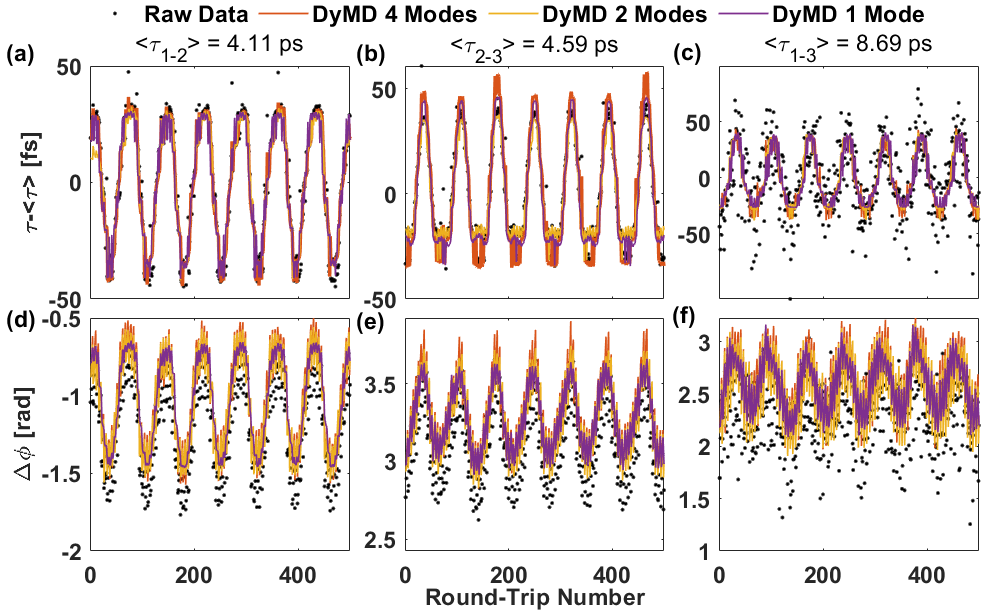}
	\caption{(a-b-c). Evolution of the inter-soliton separation for diffreent soliton pairs: $\{1-2\}$, $\{2-3\}$, and $\{1-3\}$, respectively. (d-e-f): Same as above, but for the optical phase difference between the solitons. Black points stand for data processed using a Stavitzky-Golay filter. Orange: 4-Mode DyMD. Yellow: 2-Mode DyMD. Purple: 1-Mode DyMD. }
	\label{Fig:2}
\end{figure}

\begin{figure}[!ht]
	\centering
	\includegraphics[width=\linewidth]{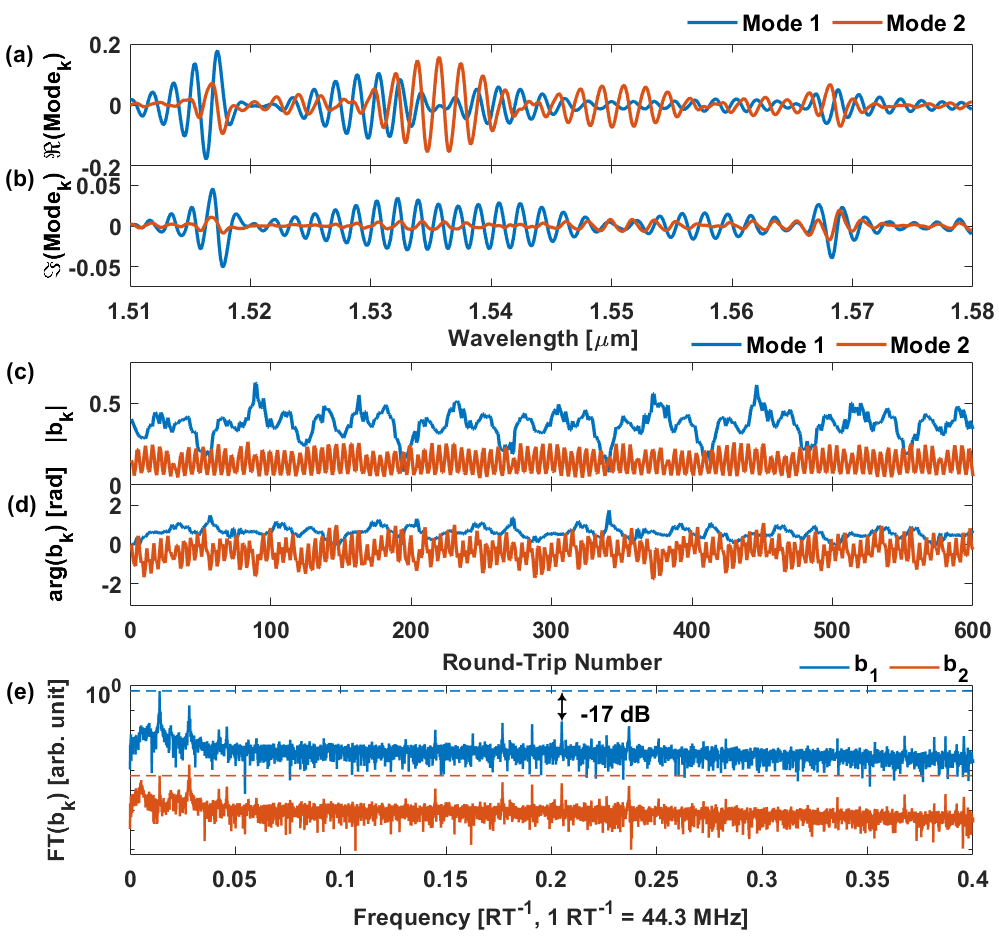}
	\caption{(a-b) Representation of the 1$^{st}$ and 2$^{nd}$ modes of the 2-Mode DyMD (c-d) Evolution of the coefficients $b_k$, $k=\{1,2\}$ used for the reconstruction. (e) Logarithmic Scale. Spectra of evolving vibration mode's amplitude $b_k$.}
	\label{Fig:3}
\end{figure}

\begin{figure}[!ht]
	\centering
	\includegraphics[width=\linewidth]{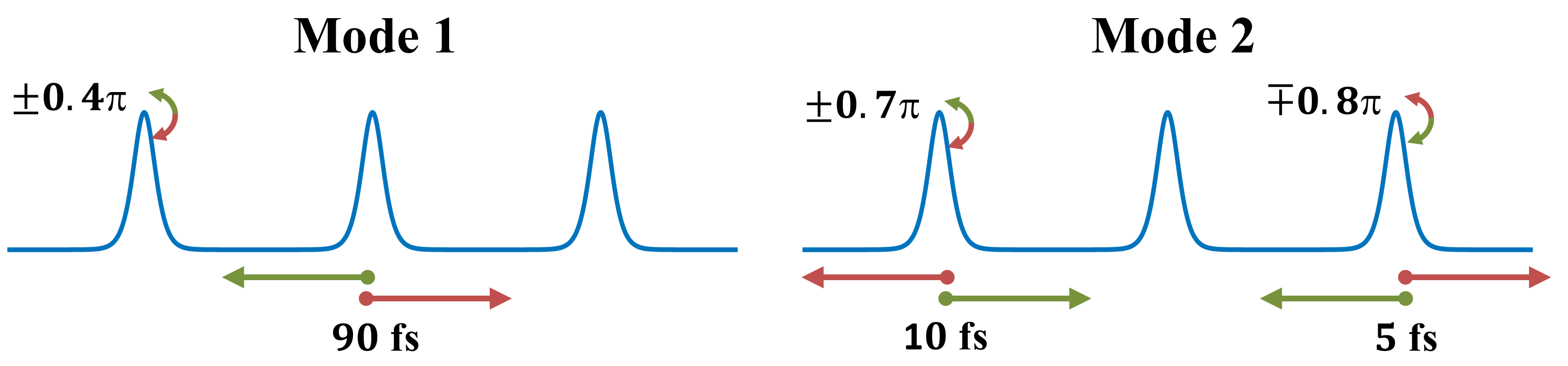}
	\caption{Schematic representation of the two vibration modes, indicating the relative displacement of the dissipative solitons (arrows, with amplitude of the displacement indicated), and the concomitant changes of optical phase. These representations can be obtained numerically by zeroing either the $b_1$ or $b_2$ coefficients during the DyMD reconstruction. For clarity, the numbers have been rounded up.}
	\label{Fig:6}
\end{figure}

This demonstration is the occasion to show that DyMD can greatly help to get a finer qualitative analysis of the dynamics of SMs, which are bound states formed by dissipative solitons \cite{Lapre2019} travelling around the laser cavity in close interaction \cite{Stratmann2005,Akhmediev1997}. The large variety of mechanisms responsible for their existence, for instance, gain recovery dynamics, cross-phase modulation effect \cite{Nimmesgern21}, laser noise \cite{Weill2016, Zhou2021}, emission of dispersive waves \cite{Soto-Crespo2003}, or acoustic phenomena \cite{Jaouen2001}, results in the SMs exhibiting numerous distinct vibration patterns . In detail, we consider here a 5-m long fiber ring laser cavity comprising 3 meters of Erbium doped fiber and a polarizer (see \fig{Fig:1}-(a)). Mode-locking is ensured by nonlinear polarization rotation followed by discrimination through a polarization beam splitter, also acting as the output coupler. After dispersion of the laser output in a -50~ps/nm Highly Dispersive Fiber (HDF), the DFT spectra are recorded by an 80~GS/s, 40~GHz electrical bandwidth, 8-bit depth oscilloscope. The photodiode has a 25~GHz bandwidth, and acts as a low pass filter for higher frequencies. This laser architecture can support the propagation of a 3-SM, characterized by the distinct interferometric fringes seen on the DFT spectra (\fig{Fig:1}-(b)). \\ As detailed later in this letter, this molecule can be described as two coupled oscillators. It exhibits in \fig{Fig:2} a complex oscillatory pattern with a principal periodicity of about 71 round-trips (RTs). Note that the SM under investigation here is not a soliton crystal because the solitons do not have the same optical phase: the leading and the trailing soliton are actually $\pi$-shifted, resulting in a more unstable interaction. Therefore, the separation between the middle an the last soliton is slightly larger that the one between the leading and middle soliton. Another consequence is that the soliton pairs have radically different properties and exhibit a very complex dynamics \cite{Tchofo19}. If the vibrations associated to the two closest pairs of solitons can be easily retrieved from the Fourier transform of the DFT spectra, the last oscillatory pattern created by the interaction of the two soliton located at each bound of the molecule clearly suffers from strong measurement noise (\fig{Fig:2}-(c,f), black dots) despite the noise level being already drastically reduced by the use of a Savitzky-Golay \cite{Savitzky1964} filter (hence a $4^{th}$ order polynomial fit over a sliding window). This type of filtering does indeed reduce the amount of noise, but it cannot suppress low frequency noise (aka. drift) and, furthermore, filter out any fast-varying data. Consequently, this signal processing helps improving the precision of the measurement to some extend, but it leaves at the same time some residual artifacts that will ultimately limit the former. In brief, the main limitation of such a filtering originates from the fact that it is conducted independently on each single shot spectrum; and therefore it does not benefit from the knowledge of the whole evolution of the single-shot spectra in order to identify truly random features, namely noise. Therefore, we performed a DyMD \cite{SCHMID_2010, Schmid2011} on the DFT dataset (\fig{Fig:1}-(b)) in order to improve the quality of the data. It consists in decomposing the instantaneous snapshot spectra $S(\lambda,n)$ at round-trip $n$ as $S(\lambda,n)=\sum_{k} \Re(~Mode_k(\lambda) b_k(n)~)$. The temporal dynamics of the laser system is then encoded in the $b_k$ coefficients (\fig{Fig:3}-(c,d), \fig{Fig:6}-(b)), whereas the $Mode_k$ contains the information related to vibration modes (\fig{Fig:3}-(a-b), \fig{Fig:6}). This decomposition method makes use of the full DFT dataset, hence it is more accurate than any local filtering and fitting technique. As the DyMD would rely on a round-trip-shifted covariance, we believe that it is very well suited for DFT data because the spectrum evolves non-chaotically round-trip after round-trip. The impact of the DyMD performed with respectively 1, 2, and 4 modes is shown in \fig{Fig:2}.

First we see in \fig{Fig:2}-(c,d) that the relative motion of the trailing and heading solitons is now retrieved correctly, whereas it suffers from very strong noise as seen for the raw data. This validates the possibility for DyMD to investigate larger soliton molecules that would otherwise remained hidden by the electronic noise (the corresponding signal is barely visible in \fig{Fig:1}-(d)). In details, we see first that the vibration patterns obtained by mean of the 4-Mode decomposition (\fig{Fig:2}, orange lines) follow almost exactly the original data. The 2-mode decomposition (yellow lines in \fig{Fig:2}) reconstructs the experimental data with an average error of 1 bit, and the maximal error never exceeds 2 bits. Indeed, if we look at \fig{Fig:1}-(b,c) where are shown the original DFT data, the reconstruction is done with little error (\fig{Fig:1}-(e)). A reconstruction error below -24~dB would actually correspond to a lossless reconstruction for a 8-bits digitization of a full scale signal. True lossless reconstruction is obtained for 8-mode decomposition (not shown here). As a side remark, the memory requirement for such a decomposition ($2^{10}$ points per spectrum, $2^{14}$ Round-Trips, 32 bits resolution) would correspond to a compression of the experimental raw data by 93\%. In contrast, the decomposition using a single mode clearly misses a few important features, like, for instance, the fast over-modulation in \fig{Fig:2}-(d).

The 2-mode DyMD appears here to be the best choice; and the corresponding modes are shown in \fig{Fig:3}-(a,b). An important point is that this decomposition does not only reduce the amount of noise, it also allows a finer interpretation of the SM vibration pattern. The $b_k$ coefficients (\fig{Fig:3}-(c,d)) are indeed completely correlated to the molecule oscillations shown in \fig{Fig:2}. We see that the weak over-modulation of 5.24 RTs periodicity that is over-imposed on the main oscillation originates almost exclusively from the second mode, whose $b_{2}$ coefficient exhibits clearly fast varying features. In contrast $b_{1}$ evolves more slowly, following a periodic pattern with high harmonic content. Detailed Fourier analysis in \fig{Fig:3}-(e) confirms this relative distribution of fast and slow oscillations between the two modes.\\ Statistically, independent oscillators are described by different modes; this is then the case here. It is actually the reason behind the failure of the single-mode DyMD as it cannot reproduce concomitantly the dynamics of two independent modes. Knowing the shape of the modes, their impact on the SM's motion can then be traced back in \fig{Fig:6}. The first vibration mode does not impact much the relative distance between the leading and trailing solitons. This fact does not change even if more modes are included in the DyMD. As a good approximation, the first mode involves the motion of the middle pulse, along with the corresponding dephasing of the first dissipative soliton. In the second mode, the middle pulse acts as an anchor, and the two other pulses move (and dephase) in an anti-symmetric manner. We can conclude that each of the two fundamental vibration patterns does not necessarily involve a single soliton pair, and that it spans over the whole molecule. Therefore, the oscillation experienced by the 3-soliton, or more complex, SM must not be simply described by looking only at the vibrations of soliton pairs. \\ If the two oscillators are independent entities, they, however, interplay with each other. Indeed both $b_1$  and $b_2$ show, respectively, modulations at high and low frequencies, which is characteristic of an interaction between the two vibration modes (\fig{Fig:3}-(e)). In detail $b_2$ encodes the fast oscillation of about 5~RTs periodicity, but exhibits at the same time a modulation at the main fundamental oscillation period (71~RTs). In \fig{Fig:3}-(e) the tone of the fast oscillation (located at about $0.2$~RT$^{-1}$) is modulated by the main slow oscillation, resulting in the creation of two equally spaced sidebands. This is the direct consequence of the slow oscillation modifying adiabatically the parameters of the fast oscillator.

\begin{figure}[!ht]
	\centering
	\includegraphics[width=\linewidth]{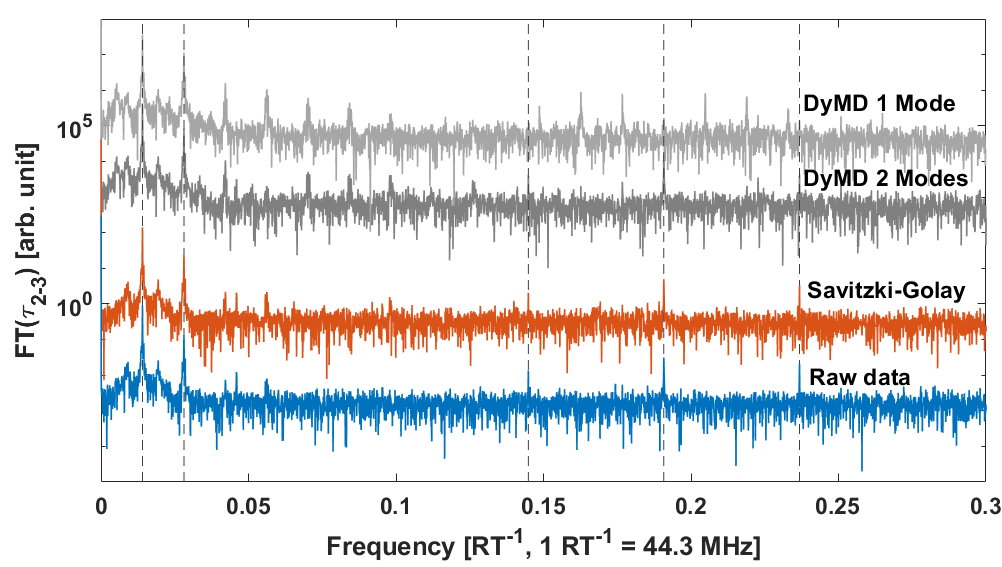}
	\caption{Spectrum of the round-trip evolution of the inter-soliton separation $\tau_{2-3}$. Blue: raw data. Red: Data processed by a 4$^{th}$ order Savitzky-Golay filter. Dark and Light Grey: Results of 2- and single-Mode DyMD, respectively. Note the apparition of spurious harmonic tone, along with the disappearance of the fast oscillation's tone around $0.2$~RT$^{-1}$.}
	\label{Fig:4}
\end{figure}

 
We would like to stress that this phenomenon can only be clearly observed because the experimental noise has been reduced so that the small fast over-modulation takes over. As a consequence, the spectral comb that characterizes the modulation of one oscillator by another is only fully revealed by the DyMD approach. To allow a global comparison are presented on \fig{Fig:4} the different spectrum of $\tau_{2-3}$ depending on the data processing. This confirms that the single mode DyMD clearly is not able to reproduce correctly the signal because it misses the fundamental tone of the fast oscillation. Compared to the raw data, which show only one, the 2-Mode DyMD exhibits up to six harmonics.

The DyMD technique also permits in-depth understanding of the molecule vibration through analysis of the $b_k$ coefficients. In literature, vibrations of SMs are usually classified depending on how the optical phase between the solitons evolves \cite{Krupa2017}, and a single oscillation pattern is usually assumed. Here the SM experiences a periodic evolution (71 RTs), set by the largest and strongest oscillator ($b_1$) that bounds the whole molecule. Still there exists a minor part of this oscillation that is caused by another oscillator, characterized by the coefficient $b_2$. Consequently the SM is under the influence of at least two limit cycles of radically different nature. The fast oscillation is of much weaker amplitude and thus it is not strong enough to take over the dynamics. The co-existence of several limit cycle attractors, and their range of attraction, is a fundamental question for a better understanding of the creation and the dissolution of SMs \cite{Grelu2012,Schreiber2007, Zavyalov2009a, Zavyalov2009}. There exists mostly scarce information regarding the stability of the molecules’motion. Assessing the latter experimentally is usually tricky as it might involve actually the destruction of the SM as soon as one tries to alter forcefully its properties. Here we see that a finer analysis of the SM motion allows to get a better estimation of its stability by simply looking at its weak fluctuations. It provides also clear indication regarding how the destabilization of the SM would occur. Considering here that the optical phase relationship between the leading and trailing solitons is not favorable for the SM’s stability, one hypothesis we could formulate is that a larger fast trembling could eventually result into the SM achieving a more stable conformation (metastability). The SM can thus switch between limit cycle attractors.

\par To conclude, we demonstrated that Dynamic Mode Decomposition is a suitable method for the analysis of DFT data. By reducing the noise, it allows the observation of minute phenomena that would have otherwise remained hidden. Moreover, the DyMD decomposition can be used to compress the data efficiently, without loss of information. Depending on the denoising which is desired, the initial data set can be sequentially reconstructed using a different number of modes. For instance, the single mode projection clearly lacks details. When it comes to the quantitative analysis of the experimental data, the DyMD is able to reveal how a complex oscillation pattern of the SM is shared between its elementary constitutive components; and separate the temporal evolution of the individual vibration modes from their spatial distribution. We are then able to describe complex vibration patterns that involve more than two pulses, and this technique could be extended to more complex solitonic systems \cite{Wu2021}.\\ As a practical example, we investigated the dynamics of a 3-Soliton molecule in an ultrafast fiber ring laser cavity. We were able to demonstrate that the soliton belonging to the same molecule may not experience the same vibration, and that at least two limit cycles of different kind could co-exist inside the same laser cavity. The optical phase difference of $\Delta\phi$ between the leading and the trailing soliton is known to create repulsion. This poses the question regarding the meta-stability of some soliton molecules configurations. And it reinforces the interest in experiments (and data analysis) of acute precision as most transition phenomena start from weak instabilities. \\ For this first proof-of-principle demonstration about the opportunities that are offered by the DyMD, we implemented the DyMD on the raw DFT spectral data, without performing any subsequent data treatment. The technique also has the denoising and compression capability. Regarding the analysis of the dynamics based on the analysis of the mode decomposition, the DyMD representation can be seen as a change of basis (or a projector, similar to a principal component analysis - PCA). Consequently, some DyMD decompositions could be more straightforward to interpret, depending on the system under study. For example, without changing the main conclusions, the DyMD could have been performed on the Fourier transformed data (\fig{Fig:1}-(d)) instead of the raw DFT data, or even on the resulting SM motion (\fig{Fig:2}). We believe that this letter paves the way for more detailed analysis of the SM dynamics. In particular, we demonstrated that the oscillatory pattern of SMs results from the interplay of individual vibration modes and we were able to extract the round-trip evolution of each mode. It is then possible to investigate further how the modes interact by constructing a simpler nonlinear model that would reproduce the interplay that is observed between the $b_k$ coefficients.

	
	\subsection*{Funding}
	This work was supported by the French ANR program, project OPTIMAL (contract ANR-20-CE30-0004)
	
	
	\subsection*{Disclosures}
	The authors declare no conflicts of interest.
	
	\subsection*{Data Availability Statement}
	Data underlying the results presented in this paper may be obtained from the authors upon reasonable request.
	

\bibliographystyle{ieeetr}
\bibliography{Biblio_PiCo}

\end{document}